\documentclass[preprint]{emulateapj}
\usepackage{booktabs}

\begin{document}

\title{Host Galaxy Properties and Black Hole Mass of Swift J164449.3+573451 from Multi-Wavelength Long-Term Monitoring and \emph{HST} Data}
\shorttitle{Swift J1644+57: Host Galaxy And Black Hole Mass}
\shortauthors{Yoon et al.}

\author{Yongmin Yoon\altaffilmark{1}, Myungshin Im\altaffilmark{1,2}, Yiseul Jeon\altaffilmark{1}, Seong-Kook Lee\altaffilmark{1}, Philip Choi\altaffilmark{3}, Neil Gehrels\altaffilmark{4}, Soojong Pak\altaffilmark{5}, Takanori Sakamoto\altaffilmark{6}, and Yuji Urata\altaffilmark{7}}

\email{yymx2@astro.snu.ac.kr, mim@astro.snu.ac.kr}

\altaffiltext{1}{Center for the Exploration of the Origin of the Universe (CEOU),
Astronomy Program, Department of Physics and Astronomy, Seoul National University, 599 Gwanak-ro, Gwanak-gu, Seoul, 151-742, Republic of Korea} 
\altaffiltext{2}{Visiting Professor, Korea Institute of Advanced Study, 85 Hoegiro, Dongdaemun-gu, Seoul 130-722, Republic of Korea
}
\altaffiltext{3}{Department of Physics and Astronomy, Pomona College, Claremont, CA 91711, USA}
\altaffiltext{4}{Astroparticle Physics Division, NASA/Goddard Space Flight Center, Greenbelt, MD 20771, USA}
\altaffiltext{5}{School of Space Research and Institute of Natural Sciences, Kyung Hee University, Yongin-si, Gyeonggi-Do 446-741, Republic of Korea}
\altaffiltext{6}{College of Science and Engineering, Aoyama Gakuin University,  5-10-1 Fuchinobe, Chuo-ku, Department of Physics and Mathematics, Sagamihara-shi Kanagawa 252-5258, Japan}
\altaffiltext{7}{Institute of Astronomy, National Central University, Chung-Li 32054, Taiwan}

\begin{abstract}
 We study the host galaxy properties of the tidal disruption object, Swift J164449.3+573451 using long-term optical to near-infrared (NIR) data. First, we decompose the galaxy surface brightness distribution and analyze the morphology of the host galaxy using high resolution \emph{HST} WFC3 images. We conclude that the host galaxy is a bulge-dominant galaxy that is well described by a single S\'{e}rsic model with S\'{e}rsic index $n=3.43\pm0.05$. Adding a disk component, the bulge to total host galaxy flux ratio (B/T) is $0.83\pm0.03$, which still indicates a bulge-dominant galaxy. Second, we estimate multi-band fluxes of the host galaxy through long-term light curves. Our long-term NIR light curves reveal the pure host galaxy fluxes $\sim500$ days after the burst. We fit spectral energy distribution (SED) models to the multi-band fluxes from the optical to NIR of the host galaxy and determine its properties. The stellar mass, the star formation rate, and the age of stellar population are $\log(M_{\star}/M_{\odot}) = 9.14^{+0.13}_{-0.10}$, $0.03^{+0.28}_{-0.03}\, M_{\odot}$/yr, and $0.63^{+0.95}_{-0.43}$ Gyr. Finally, we estimate the mass of the central super massive black hole which is responsible for the tidal disruption event. The black hole mass is estimated to be $10^{6.7\pm0.4}\, M_{\odot}$ from $M_{\mathrm{BH}}$ -- $M_{\star,\mathrm{bul}}$ and $M_{\mathrm{BH}}$ -- $L_{\mathrm{bul}}$ relations for the $K$ band, although a smaller value of $\sim10^5\, M_{\odot}$ cannot be excluded convincingly if the host galaxy harbors a pseudobulge. \\
\end{abstract}


\section{Introduction} \label{sec:Intro}
 \emph{Swift} J164449.3+573451 (hereafter, Swift J1644+57) was first discovered by the \emph{Swift} Burst Alert Telescope (BAT) at 12:57:45 UT on 28 March 2011 \citep{Burrows2011,Levan2011}. Some evidences suggest that Swift J1644+57 is a tidal disruption of a star by a supermassive black hole (SMBH). This phenomenon triggered the BAT three times after the initial trigger during the first few days \citep{Burrows2011, Levan2011}. The late-time X-ray light curve was extended to a longer period by following the expected power-law decay for the tidal disruption of a star, i.e., $t^{-5/3}$ \citep[e.g.,][]{Rees1988}. Finally, the source of X-ray, IR, and radio emissions were well matched up with the center of the host galaxy where a SMBH resides \citep{Levan2011,Zauderer2011}. 

 There have been many studies performed to understand the nature of this event, such as the characteristics of the star that was disrupted. Such a question is closely connected to the SMBH mass ($M_{\rm BH}$). The disruption of a solar-type star is possible for all  $M_{\rm BH} < 10^8\, M_{\odot}$ \citep{Rees1988,Cannizzo1990,Bloom2011}, but compact stars like a white dwarf can be disrupted only if $M_{\rm BH} < 10^{5}\, M_{\odot}$ \citep{kp2011}. If so, then this kind of event provides the interesting possibility of discovering intermediate-mass black holes.

 Unfortunately, there has been controversy concerning the mass of the SMBH. \citet{Burrows2011} provided a rough estimate of the SMBH mass of $\sim2\times 10^{7}\,M_{\odot}$ using a black hole mass -- luminosity relation and the lower limit of $\sim10^6\,M_{\odot}$ based on the X-ray variability. Similarly, \citet{Levan2011} estimated $M_{\rm BH}$ to be $2\times 10^{6}$ -- $10^{7}\,M_{\odot}$, derived from $K$-band luminosity, but at that time, $K$-band luminosity contained a significant amount of the transient light. \citet{mg2011} utilized a relation between the black hole mass, the radio luminosity, and the X-ray luminosity, and found $M_{\rm BH}$ $\sim10^{5.5}\,M_{\odot}$. Using a quasi-periodic oscillation resonance hypothesis, \citet{al2012} provided an $M_{\rm BH}$ estimate of $\sim10^{5}\,M_{\odot}$. \citet{kp2011} concluded that a white dwarf was tidally disrupted and the mass of SMBH is less than $10^{5}\,M_{\odot}$ in light of the short timescales of the X-ray light curve. In summary, the $M_{\rm BH}$ estimates have centered around the two discrepant values of $10^{7}\,M_{\odot}$ and $10^{5}\,M_{\odot}$ or less. A better understanding of the host galaxy properties is needed to clear up the situation. 

\begin{deluxetable*}{cccccc}
\tabletypesize{\scriptsize}
\tablewidth{0pt}
\tablecaption{\emph{HST} WFC3 Data Log} 
\tablehead{
\colhead{Observation date (UT)} & \colhead{MJD\tablenotemark{a}} & \colhead{Days since trigger} & \colhead{band} & \colhead{Exptime (s)} & \colhead{Magnitude (AB)}
}
\startdata \\
2011-04-04 & 55655.147614 & 6.6 & F606W & 1260 & 22.69$\pm$0.01\\
2011-08-04 & 55777.276876 & 129 & F606W & 4160 & 22.76$\pm$0.01\\
2011-12-02 & 55897.684390 & 249 & F606W & 1113 & 22.77$\pm$0.01\\
2013-04-12 & 56394.429204 & 746 & F606W & 2600 & 22.74$\pm$0.01\\
\\
\tableline \\
2011-04-04 & 55655.132654 & 6.6 & F160W & 997 & 20.68$\pm$0.01\\
2011-08-04 & 55777.257148 & 129 & F160W & 1412 & 21.09$\pm$0.01\\
2011-12-02 & 55897.702220 & 249 & F160W & 1209 & 21.22$\pm$0.02\\
2013-04-12 & 56394.295795  & 746 & F160W & 2812 & 21.55$\pm$0.02\\
\enddata
\tablenotetext{a}{Exposure start time in Modified Julian Date (MJD)}
\label{HSTtab}
\end{deluxetable*}

\begin{deluxetable*}{cccccc}
\tabletypesize{\scriptsize}
\tablewidth{0pt}
\tablecaption{\emph{Spitzer} IRAC Data Log} 
\tablehead{
\colhead{Observation date (UT)} & \colhead{MJD\tablenotemark{a}} & \colhead{Days since trigger} & \colhead{band} & \colhead{Exptime (s)} & \colhead{Magnitude (AB)}
}
\startdata \\
2011-04-28 & 55679.975316 & 31.4 & 3.6$\mu$m & 1250 & 19.50$\pm$0.02 \\ 
2011-10-31 & 55865.023052 & 216 & 3.6$\mu$m & 1253 & 21.49$\pm$0.06 \\ 
2012-02-24 & 55981.541644 & 333 & 3.6$\mu$m & 1252 & 21.70$\pm$0.07 \\ 
\\
\tableline \\
2011-04-28 & 55679.975316 & 31.4 & 4.5$\mu$m & 1250 & 19.30$\pm$0.01 \\ 
2011-10-31 & 55865.023052 & 216 & 4.5$\mu$m & 1253 & 21.25$\pm$0.05 \\ 
2012-02-24 & 55981.541644 & 333 & 4.5$\mu$m & 1252 & 21.57$\pm$0.08 \\ 
\enddata
\tablenotetext{a}{MJD in UTC at data collection event (DCE) start}
\label{spitzertab}
\end{deluxetable*}

 In order to more accurately estimate the SMBH mass and better constrain the properties of the host galaxy, we analyze the morphology and the surface brightness profile of the host galxaxy based on high-resolution \emph{Hubble Space Telescope} (\emph{HST}) images and estimate the multi-band fluxes of the host galaxy using our long-term monitoring data lasting more than 2.4 years. We fit the multi-band spectral energy distribution (SED) of the host galaxy luminosity with stellar population synthesis models, and then obtain the properties of the galaxy. Finally, we provide our best estimate of $M_{\rm BH}$ based on the host galaxy properties.

 This is the second of a series of two papers. In the first paper (M. Im et al. 2015 in preparation, hereafter, Im15), we present the dataset of the long-term monitoring campaign and an analysis of the late-time light curve. 

Throughout the paper, we selected \emph{H$_0=70$}km s$^{-1}$Mpc$^{-1}$, $\Omega_{\Lambda}=0.7$, and $\Omega_{m}=0.3$ as cosmological parameters and adopt the AB magnitude system.
\\

\section{Observations and Data}
We observed  \emph{Swift} J1644+57 using Wide Field Camera (WFCAM) on United Kingdom Infrared Telescope (UKIRT) for nearly 2.4 years following the burst as a part of our gamma-ray burst (GRB) and transient observation program \citep{Lee2010}. We observed intensively in the $K$ band among $Z, Y, J ,H $, and $K$ bands of WFCAM. The number of epochs of $K$ band data used for thie analysis is $101$ and the last data were observed at $884.7$ days after the initial BAT trigger. The numbers of epochs for the $Y,J,H$-band data are $3,15,28$ and the last data were observed at $\Delta t = 712.1$, $710.1$, and $884.7$ days, respectively, where $\Delta t$ is the number of days since the initial BAT trigger. We have only one epoch of data for the UKIRT $Z$ band which was observed at $\Delta t = 723.0$ days. 

 We also observed \emph{Swift} J1644+57 using Camera for QUasars in EArly uNiverse \citep[CQUEAN;][]{KimE2011,Park2012,Lim2013} on the 2.1m Otto-Struve telescope of the McDonald Observatory in $g, r, i, z$, and $Y$ bands. The numbers of epochs for the $g, r, i, z,$ and $Y$ band data are $2,2,14,14$, and $2$ and the last data were observed at $\Delta t = 25.7, 217.5, 526.6, 526.6$, and $25.8$ days respectively. The UKIRT and CQUEAN observation logs and photometry results are described in Im15. 

 In addition, we also used data from \citet{Burrows2011} and \citet{Levan2011} for the earlier optical and near-infrared (NIR) data. 

 Morphology analysis requires high-resolution images because this object is so compact that it is virtually a point source in the UKIRT and CQUEAN images. For the high-resolution images, we obtained \emph{HST} WFC3 multi-drizzled, stacked images\footnote{Based on observations made with the NASA/ESA Hubble Space Telescope, obtained from the data archive at the Space Telescope Science Institute. STScI is operated by the Association of Universities for Research in Astronomy, Inc. under NASA contract NAS 5-26555.} available in the MAST database. We used  F606W-, F160W-band data and the number of epochs in each two band is four. These data were observed at $\Delta t = 6.6, 129, 249,$ and $746$ days. The \emph{HST} WFC3 data are summarized in Table~\ref{HSTtab}.

\begin{figure*}[!t]
\centering
\includegraphics[scale=0.225,angle=00]{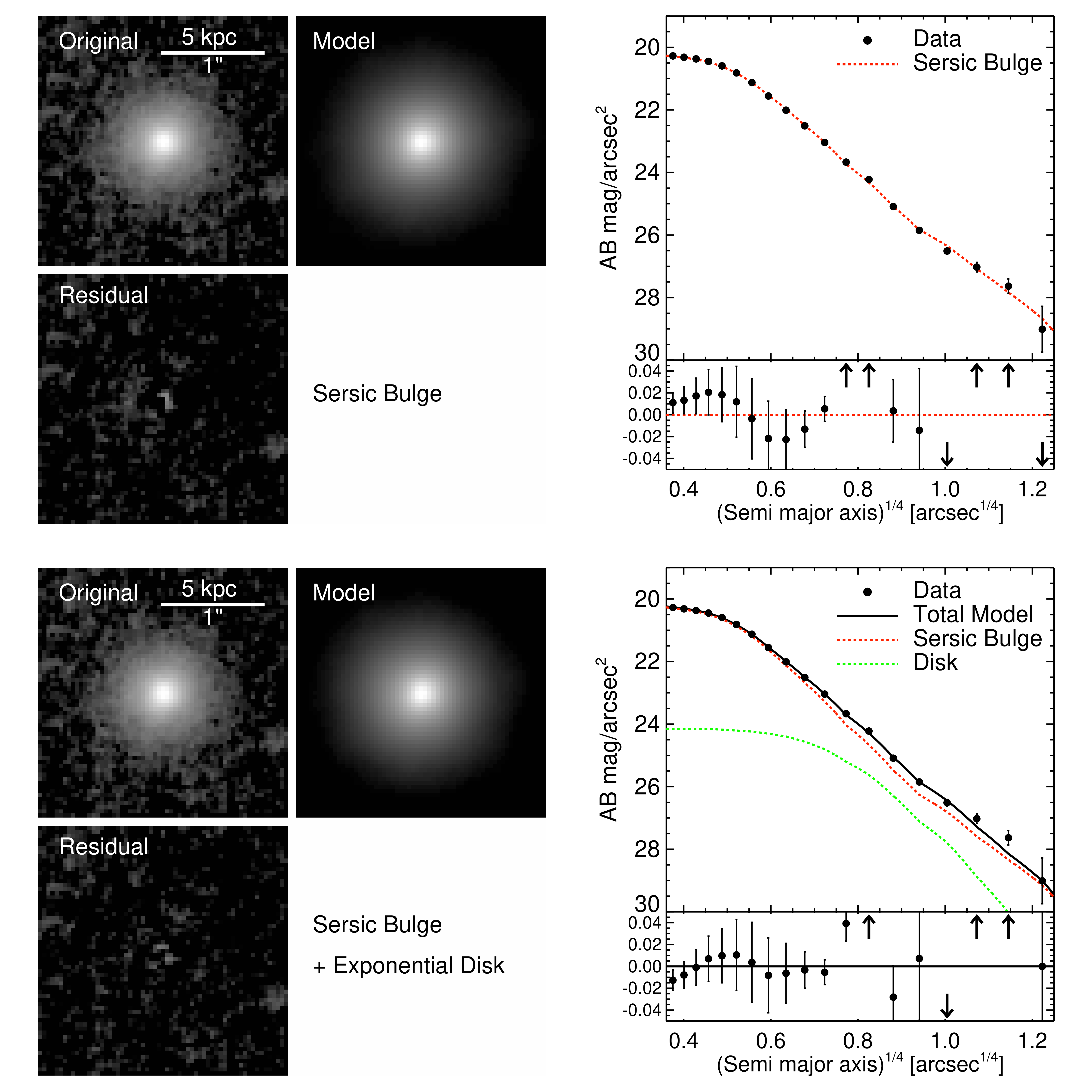}
	\caption{Images of the host galaxy in F606W, the best-fit two-dimensional models from GALFIT, and the residuals for the single S\'{e}rsic component model and the S\'{e}rsic bulge $+$ exponential disk model. Right panels show one-dimensional profiles of the host galaxy and each model component.
		\label{sbfig}}
\end{figure*}
\begin{figure*}[!t]
\centering
\includegraphics[scale=0.225,angle=00]{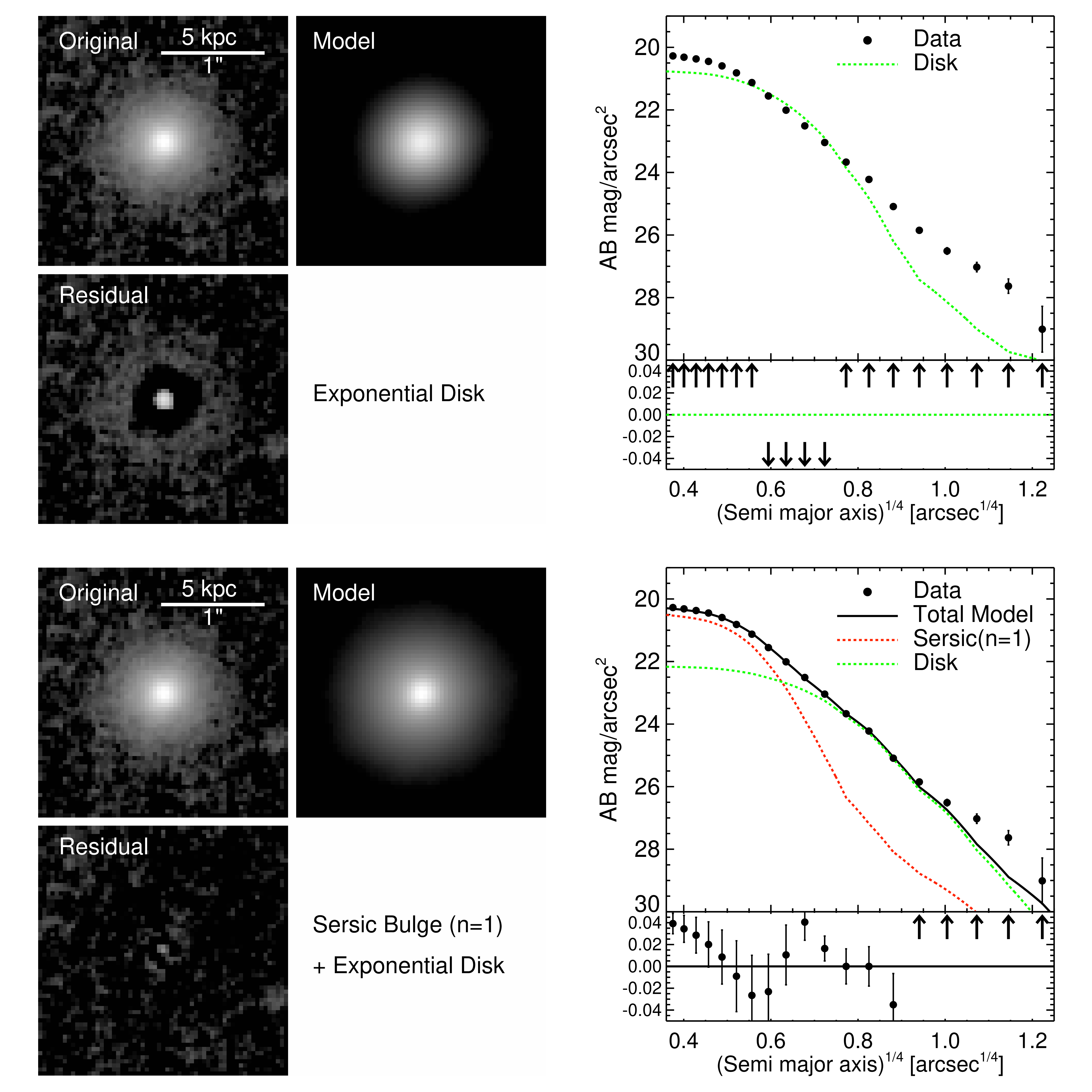}
	\caption{Same as Figure~\ref{sbfig}, but for the single exponential disk model and the double exponential profile model. 
		\label{sb2fig}}
\end{figure*}

\begin{figure}[!t]
\centering
\includegraphics[scale=0.22,angle=00]{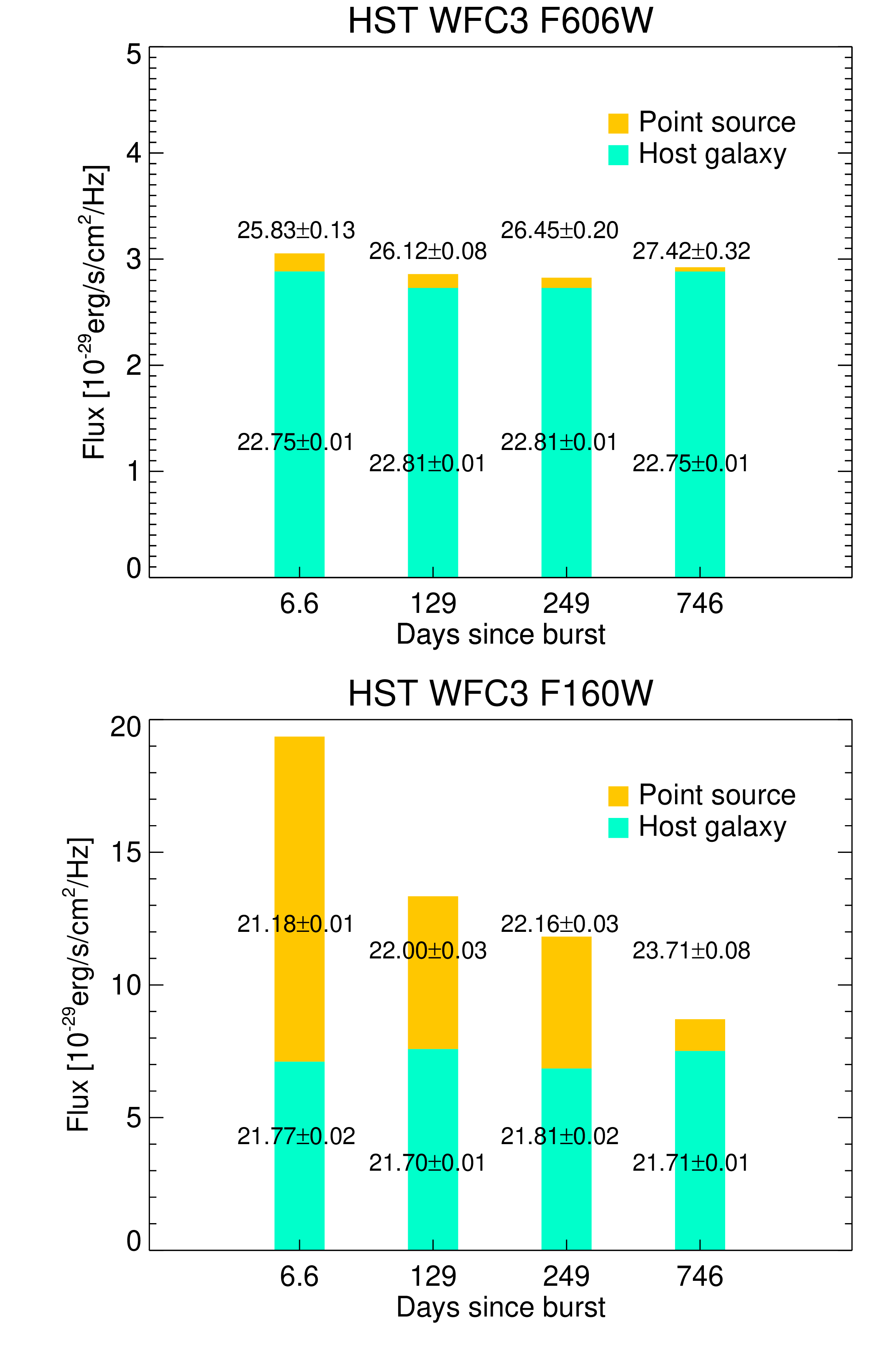}
	\caption{Flux fractions of the GALFIT models as a function of time. We used the model which consists of a single S\'{e}rsic bulge with $n=3.43$ and a point source component. The upper panel shows the results for the F606W-band images, while the lower panel shows the results for the F160W-band images. Magnitudes of the model components are also shown.
		\label{hstfig}}
\end{figure}

 To supplement the NIR observation data, we used the \emph{Spitzer} IRAC 3.6, 4.5$\mu$m post basic calibrated data (PBCD) from the NASA/IPAC Infrared Science Archive. These were observed at $\Delta t = 31.4, 216.5$, and $333.0$ days. A log of the \emph{Spitzer} IRAC 3.6, 4.5$\mu$m data are shown in Table~\ref{spitzertab}.

The flux measurements were performed by SExtractor software\footnote{We used aperture magnitudes with aperture correction.} \citep{Bertin1996} except for the \emph{HST} images for which GALFIT \citep{Peng2010} models were used for the flux measurements. 

 For the X-ray data, we used \emph{Swift}/XRT data taken from the \emph{Swift} archive and \emph{XMM-Newton} data\footnote{Based on observations obtained with XMM-Newton, an ESA science mission with instruments and contributions directly funded by ESA Member States and NASA} from the \emph{XMM-Newton} Science Archive.\\

\section{Morphology of the host galaxy}  \label{sec:Mor}
 We analyzed the surface brightness profile of the host galaxy, in order to determine the bulge fraction and its nature. We used the \emph{HST} images and GALFIT software to fit two-dimensional models to the light distribution of the host galaxy. To construct the point spread function (PSF), we selected $\sim 5$ isolated stars with signal-to-noise ratios $\gtrsim 300$ in the vicinity of \emph{Swift} J1644+57, and co-added them.

 We used error images that are created by GALFIT for the fitting. For GALFIT to create the error image properly, we modified the unit of ADU and the image header values such that $\mathrm{GAIN} \times \mathrm{ADU} \times  \mathrm{NCOMBINE} =  \mathrm{[electrons]}$ as recommended in GALFIT website\footnote{http://users.obs.carnegiescience.edu/peng/work/galfit/TOP10.html}.

 A crucial factor affecting the fitting results is background subtraction. For the background determination, we set 6 annuli with the radii logarithmically increasing between 2.5 and 9 times the radius of an ellipse for which pixel values are $1.5\sigma$ of the background noise. We centered the annuli on the center of the host galaxy, set the minimum width of the annuli to be $\sim1.3$ arcsec ($\sim33$ pixel) for F606W images and $\sim2$ arcsec ($\sim16$ pixel) for F160W images, and augmented the widths in step with the logarithmically growing radii. We then derived the mean pixel values of each annulus. Finally, we adopted their mean value as the background value.

 Our surface brightness fit was carried out using a deep, stacked image of the data taken with F606W at $\Delta t = $ 129, 249, and 749 days. It has been known that the transient component is negligible in the optical bands bluer than $i$ even at the early time \citep{Burrows2011, Levan2011}. The use of the stacked, late-time image in the F606W band makes the transient component more negligible.  On the other hand, NIR-bands, including F160W (similar to $H$ band of WFCAM), are known to contain a significant transient component which may affect the host galaxy analysis. Furthermore, the spatial resolution of the F606W images is better by a factor of three than that of F160W, which greatly helps the surface brightness fitting. The other \emph{HST} data were also analyzed to understand the importance of the transient component, and the results for the transient component are presented later in this section.

 For the galaxy models, we used the S\'{e}rsic \citep{Sersic1968}, de Vaucouleurs \citep{deVa1948}, and exponential disk profiles or a combination of thereof. The S\'{e}rsic profile is described as
\begin{displaymath}
        \Sigma(r)=\Sigma_{e}\mathrm{exp}\Bigg[-\kappa\Bigg(\bigg(\frac{r}{r_e}\bigg)^{1/n}-1\Bigg)\Bigg],
\end{displaymath}
where $\Sigma_{e}$ is the surface brightness at the effective radius $r_e$, and $n$ is the S\'{e}rsic index. $\kappa$ is a variable parameter denpendent on $n$, where $n=4$ and $1$ correspond to the de Vaucouleurs and exponential profiles, respectively. Although $n=4$ is commonly quoted for the ellipticals and classical bulges, the S\'{e}rsic index of ellipticals and classical bulges can assume a value in the range $2\lesssim n \lesssim6$, whereas pseudobulges have $n\lesssim 2$ \citep{Fisher2008, Fisher2010}.

 All of the model parameters such as ellipticity and center positions of the different components, were set free in the fitting procedure. 

 Figures~\ref{sbfig} and \ref{sb2fig} show images of the host galaxy, the two-dimensional models, and the residuals (i.e., the model subtracted images), for four different models: (1) a single S\'{e}rsic; (2) a  S\'{e}rsic bulge + exponential disk; (3) an exponential disk; and (4) a double exponential profile models. The figures also show one-dimensional profiles (along the major axis) of the host galaxy and those of each model component, which are converted through the IRAF\footnote{IRAF is distributed by the National Optical Astronomy Observatory, which is operated by the Association of Universities for Research in Astronomy (AURA) under a cooperative agreement with the National Science Foundation.} ELLIPSE task. In addition to the profiles, the differences between the data and the model profiles are shown. The results of each fit are summarized in Table~\ref{galfittab}. Both the single S\'{e}rsic model with $n=3.43\pm0.05$ and the S\'{e}rsic bulge with $n=3.39\pm0.11$ $+$ exponential disk model fit the data well ($\chi_{\nu}$ $\sim 1.2$ -- $1.3$).  When the disk component is added, the bulge to total host galaxy flux ratio (B/T) is $0.83\pm0.03$. On the other hand, the single exponential disk model provides a poor fit to the data as shown in Figure 2 and with $\chi_{\nu}=6.54$. The double exponential profile model fits the data nearly as well as the single S\'{e}rsic model and the S\'{e}rsic bulge+disk model in terms of $\chi_{\nu}^{2}$. However, the analysis of the one-dimensional surface brightness profile shows that the model does not follow the outer part of the profile well, demonstrating a relatively steeper decline than that of the single S\'{e}rsic model and the S\'{e}rsic bulge $+$ exponential disk model. This model gives B/T $= 0.36$, suggesting a significant bulge component. Therefore, we conclude that the host galaxy of \emph{Swift} J1644+57 is bulge-dominant. We also conclude that the bulge is likely to have a S\'{e}rsic index higher than $3$ regardless of the existence of the disk. This value corresponds to the range of the classical bulges \citep{Fisher2008, Fisher2010}. We cannot completely exclude the case where the bulge is pseudobulge with $n \sim 1$, but even in this case, the object has a significant bulge.

\begin{deluxetable*}{ccccccccccc}
\tabletypesize{\scriptsize}
\tablewidth{0pt}
\tablecaption{Surface Brightness Fitting Result} 
\tablehead{
\colhead{} & \multicolumn{3}{c}{Bulge} & \colhead{} & \multicolumn{2}{c}{Disk}& \colhead{} & \colhead{} & \colhead{} & \colhead{} \\ 
\cline{2-4}\cline{6-7}\\
\colhead{Galaxy model} &  \colhead{$m_{\mathrm{b}}$ [AB]} & \colhead{$n$} & \colhead{$r_{\mathrm{eff}}[\mathrm{kpc}]$} & & \colhead{$m_{\mathrm{d}}$ [AB]} &  \colhead{$r_{\mathrm{s}}[\mathrm{kpc}]$} & \colhead{} & \colhead {B/T} & \colhead {$m_{\mathrm{t}}$ [AB]} & \colhead{$\chi_{\nu}^{2}$}\\ 
\colhead{(1)} & \colhead{(2)} & \colhead{(3)} & \colhead{(4)} & \colhead{} & \colhead{(5)} & \colhead{(6)} & \colhead{} & \colhead{(7)} & \colhead{(8)} & \colhead{(9)}
}
\startdata
B & $22.77\pm0.01$ & $3.43\pm0.05$ & $1.01\pm0.01$ & & \nodata & \nodata & &  \nodata & $22.77\pm0.01$ & 1.322\\
B$+$D & $23.01\pm0.03$ & $3.39\pm0.11$ & $0.79\pm0.03$ & & $24.75\pm0.14$ & $1.21\pm0.07$ & & $0.83\pm0.03$ & $22.81\pm0.03$ & 1.223\\
\tableline \\
B$(n=4)$ & $22.72\pm0.00$ &4 (fixed) & $1.09\pm0.01$ & & \nodata & \nodata & & \nodata & $22.72\pm0.00$ & 1.275\\
D & \nodata & \nodata & \nodata & & $23.07\pm0.00$ & $0.47\pm0.00$ & & \nodata & $23.07\pm0.00$ & 6.547\\
B$(n=4)$$+$D & $23.01\pm0.03$ & 4 (fixed) & $0.85\pm0.02$ & & $24.64\pm0.10$ & $0.95\pm0.03$ & & $0.82\pm0.03$ & $22.79\pm0.03$ & 1.228\\
S$(n=1)$$+$D & $23.99\pm0.01$ & 1 (fixed) &  $0.30\pm0.00$ & & $23.35\pm0.00$ & $0.92\pm0.01$ & & $0.36\pm0.00$ & $22.87\pm0.00$ & 1.351
\enddata
\tablecomments{Column~1:  galaxy model for the two-dimensional fitting. B: S\'{e}rsic bulge, D: exponential disk , B$(n=4)$: de Vaucouleurs bulge. S$(n=1)$: S\'{e}rsic profile with fixed $n=1$ (exponential profile). Column~2: AB magnitude of the bulge component. Column~3: S\'{e}rsic index for the bulge model. Column~4: effective radius. Column~5: AB magnitude of the disk component. Column~6: scale length of the disk component. Column~7: bulge to total light ratio. Column~8: total magnitude. Column~9: reduced $\chi^2$ for the fitting model defined as
\begin{displaymath}
        \chi_{\nu}^{2}=\frac{1}{N_\mathrm{DOF}} \sum_{x=1}^{nx}\sum_{y=1}^{ny}\frac{(f_\mathrm{data}(x,y)-f_\mathrm{model}(x,y))^2}{\sigma(x,y)^2},
\end{displaymath} 
where $f_\mathrm{data}(x,y)$ and $f_\mathrm{model}(x,y)$ mean the input data and the model images, respectively. $N_\mathrm{DOF}$ is the degree of freedom. $\sigma(x,y)$ is the error image. Here, sum is only over all $nx$ and $ny$ pixels satisfying $1.5\sigma$ of the background noise.}
\label{galfittab}
\end{deluxetable*}

\begin{figure*}[!t]
\centering
\includegraphics[scale=0.36,angle=00]{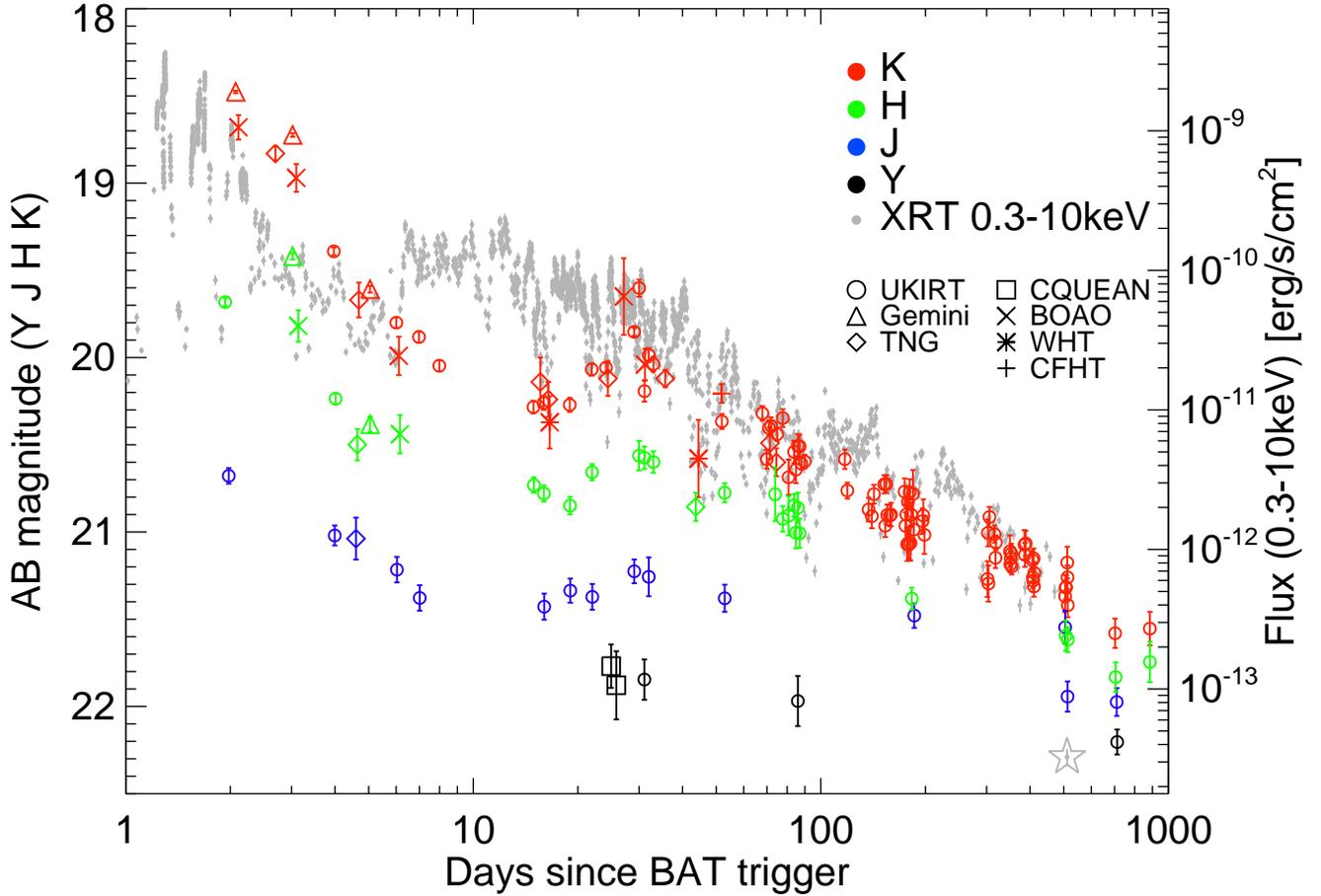}
	\caption{NIR and the X-ray light curves of the \emph{Swift} J1644+57.
                      The light curves of $Y, J ,H ,K$ bands with the early data from \citet{Burrows2011}, \citet{Levan2011}, and \emph{Swift}/XRT $0.3$ -- $10$keV are shown. All of the light curves have similar shapes except that the X-ray light curve is $\sim$15 days ahead of the NIR light curves. After $\sim$500 days the X-ray emission was rapidly declined as shown in with star mark for the last X-ray data from the \emph{Swift}/XRT.
                  \label{ukirtfig}}
\end{figure*}

\begin{figure}[!t]
\centering
\includegraphics[scale=0.22,angle=00]{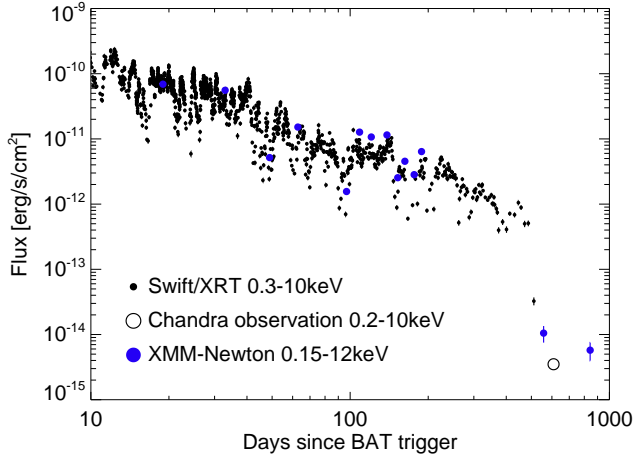}
	\caption{\emph{Swift}/XRT and \emph{XMM-Newton} data as well as recent \emph{Chandra} observation of \emph{Swift} J1644+57. The X-ray flux abruptly declined after $\sim$500 days since the BAT trigger. 
		\label{xfig}}
\end{figure}

\begin{figure}[!t]
\centering
\includegraphics[scale=0.175,angle=00]{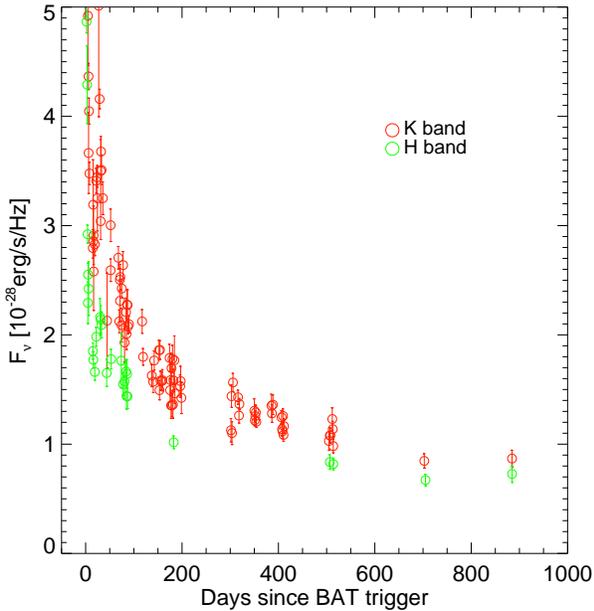}
	\caption{$H$- and $K$-band light curves in linear scale. Some of the data points at very early times are cut in order to highlight the late-time light curves. The magnitudes of the latest $H, K$ bands converge to single values, suggesting that the transient component has disappeared at $\Delta t > 500$ days.
		\label{linfig}}
\end{figure}

\begin{figure*}[!t]
\centering
\includegraphics[scale=0.30,angle=00]{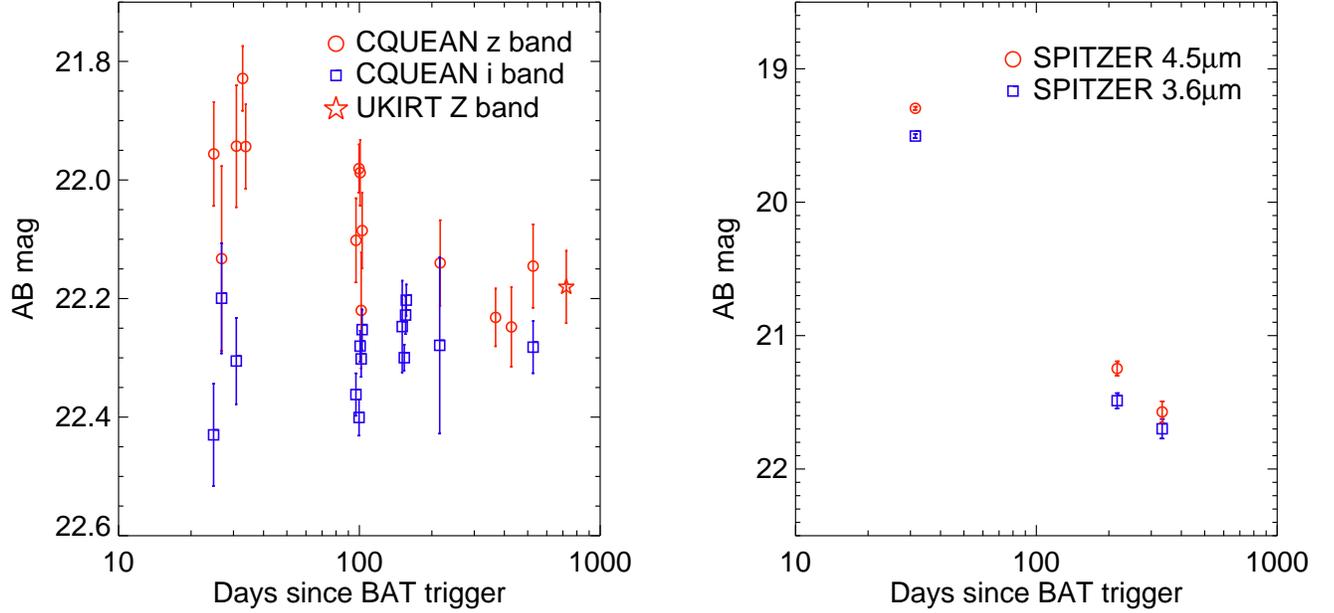}
	\caption{Light curves of CQUEAN $i$ and $z$ bands, UKIRT $Z$-band data, and \emph{Spitzer} IRAC 3.6$\mu$m and 4.5$\mu$m bands. Note that the $z$-band flux decreases with time, while the $i$-band flux is almost constant with time. Considerable magnitude changes in the \emph{Spitzer} IRAC 3.6$\mu$m and 4.5$\mu$m bands can be seen in the right plot.
		\label{lcfig}}
\end{figure*}

\begin{figure}[!t]
\centering
\includegraphics[scale=0.175,angle=00]{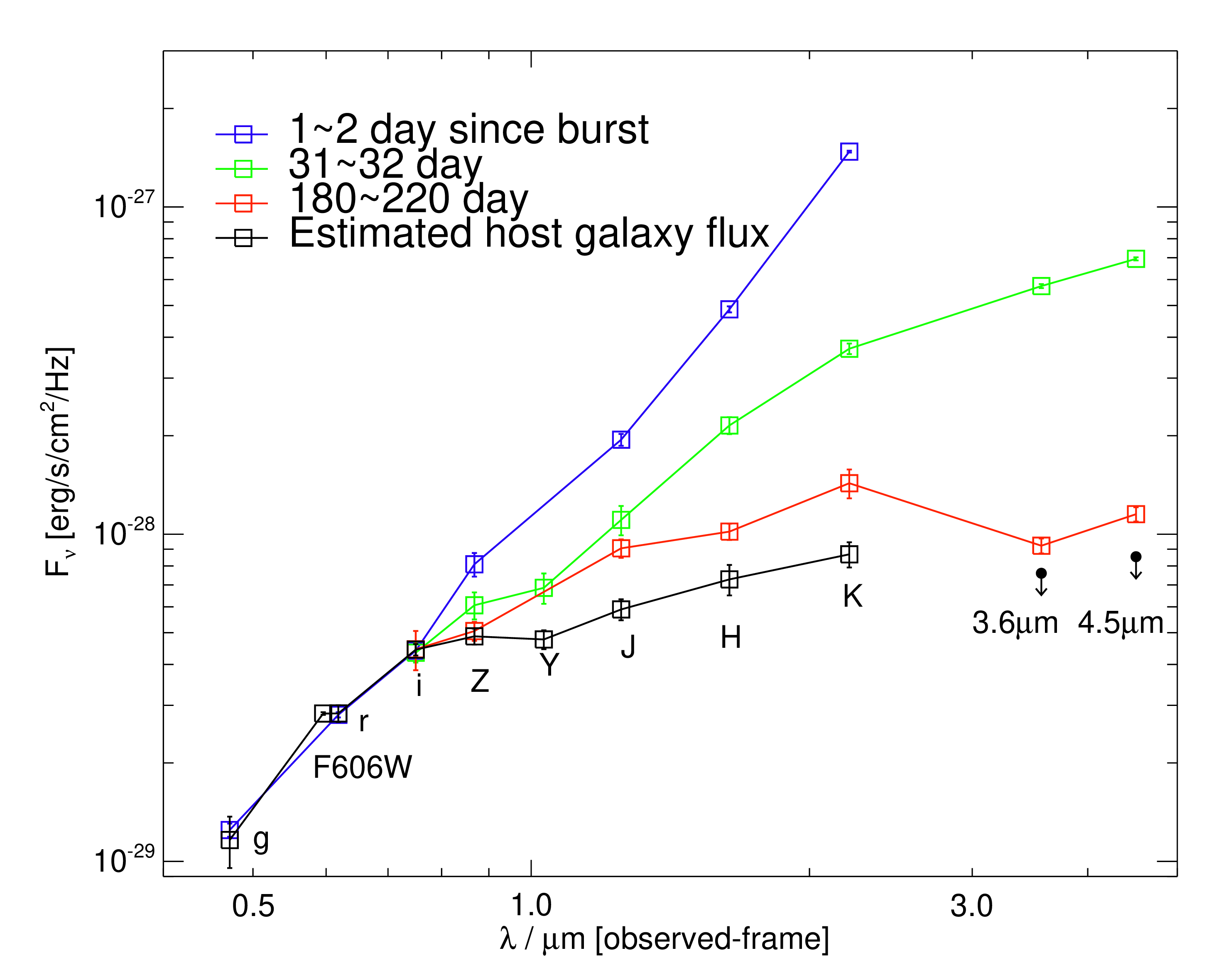}
	\caption{Temporal change of the SED of \emph{Swift} J1644+57. The fluxes in the redder bands show the substantial changes with time, whereas the bluer-band fluxes do not vary with time.
		\label{changefig}}
\end{figure}

Additionally we fit the observed surface brightness profile with a single de Vaucouleurs bulge model and a de Vaucouleurs bulge $+$ exponential disk model. The results of these fits are nearly identical to that of the single S\'{e}rsic and the S\'{e}rsic bulge+disk models.

 To estimate the transient component flux, we fit all the F606W and F160W images with a model containing both the point source (transient) and the host galaxy components. Here, we adopt a single S\'{e}rsic profile with a fixed S\'{e}rsic index ($n=3.43$) for the host galaxy component, and a PSF profile for the transient component. The compactness of the host galaxy and the bright transient component in the F160W images create a serious degeneracy, particularly between the effective radius and the S\'{e}rsic index, when fitting multi-component models. To alleviate the degeneracy, we fixed the S\'{e}rsic index to be $n=3.43$, similar to that of the F606W band. The flux fractions of the models as a function of time are shown in Figure~\ref{hstfig}. In the case of F606W, the flux fraction from the transient component is very small or nonexistent, while the transient component is very bright in the earliest F160W band, even brighter than the entire host galaxy. The result reflects a very red color to the transient well, and justifies the exclusion of the point source component in the late-time F606W images during the host galaxy analysis. The point source contribution declines rapidly as time goes on in F160W, but it contributes to the total flux of the object until around $\Delta t = 750$ days. On the other hand, the fluxes of the host galaxy component are almost constant in both bands over the entire period. The magnitude of host galaxy in the F160W band is $\sim21.75$ mag, and as we shall see in the next section, this is the same as for the last data point of the UKIRT $H$-band light curve, suggesting that the flux of the last data point in the NIR light curve represents the host galaxy flux. 
\\

\section{Light Curves}  \label{sec:LC}
 In this section, we show long-term observation results and estimate multi-band fluxes of the host galaxy of \emph{Swift} J1644+57. Figure~\ref{ukirtfig} shows the  $Y, J, H$, and $K$ light curves. The gray data points in the background show \emph{Swift}/XRT $0.3$ -- $10$keV data. The $J,H,K$ light curves resemble each other. The NIR light curves rapidly decline until $\Delta t \simeq 10$ days, turn up again with a second peak at $\Delta t \simeq 30$ days, and decline again steadily. The behaviors of these NIR light curves are very similar to that of the X-ray light curve except that the X-ray light curve appears shifted ahead of the NIR light curves at a time of $\sim$15 days. The similar shapes of these light curves indicate that the origins of the X-ray emission and NIR emission are related to each other. On the other hand, the time gap between these two emissons denotes that the X-ray source and NIR source are separated from each other as much as the time gap. \citet{Bloom2011} suggested that X-ray source is in the close vicinity of the black hole due to the fact that the X-ray emission shows very rapid, high variability, while the IR and the radio sources are located a large distance from the black hole on account of the relatively smooth and small variability. They argued that the jet generated by black hole collides with the surrounding medium where the electrons are accelerated by the jet. These high-speed electrons emit the IR to radio photons through synchrotron radiation.

 The jet seems to be nearly turned off at $\Delta t \simeq 500$ days in light of the fact that there is an abrupt decrease in flux of a factor of $\sim10$  or more, which can be seen in all the \emph{Swift}/XRT, \emph{Chandra}, and \emph{XMM-Newton} data \citep{Levan2012, Zauderer2013}. This late-stage turn-off of X-ray flux is also shown in Figure~\ref{xfig}. If the jet was turned off, then the transient components of the NIR fluxes must be quenched following the X-ray flux, and it is expected that the fluxes of the pure host galaxy of \emph{Swift} J1644+57 were revealed at that time. As we can see in Figure~\ref{linfig}, which shows the $H$-, $K$- bands light curves in linear scale, the fluxes of the $H, K$ bands converge to single values at late-time. Furthermore, the latest $H$-band magnitude is nearly the same as that of the host galaxy of the F160W-band images, shown in the results of the model fitting in \S\ref{sec:Mor}. This evidence indicates that it is reasonable to regard the NIR ($Y, J, H$, and $K$ band) fluxes of the last data, taken at $\Delta t = \sim 700$ or $884$ days, as those of the pure host galaxy.

 The left panel of Figure~\ref{lcfig} shows the light curves of the CQUEAN $i$- and $z$-band and the UKIRT $Z$-band data. In the case of the $z$ band, the fluxes from the object slightly decrease with time. We also regard the flux of the last data, that is the UKIRT $Z$-band data, as the $Z$-band flux of the host galaxy since it is observed far beyond expected quenching time of jet. On the other hand, $i$-band fluxes are virtually constant, demonstrating that the transient components are basically non-existent in the $i$ or bluer bands \citep[Figure~\ref{hstfig};][]{Burrows2011, Levan2011}. We take the flux of the last data of the $i$ band as that of the host galaxy. The number of CQUEAN $g$- and $r$-band data points are scarce compared to the NIR data and the last data were observed at early time ($\Delta t =25.7, 217.5$ days, respectively). However, there is little or no change between the very early-time magnitudes from \citet{Levan2011} and our $g$- and $r$-band magnitudes in the same way as the fluxes of the $i$ and F606W bands. Therefore, we consider the $g$- and $r$-band fluxes of the last epoch data as those from the host galaxy. Furthermore, we also consider the magnitude of the single S\'{e}rsic model of the stacked \emph{HST} WFC3 F606W image as that of the host galaxy since the point source contribution to whole flux is negligible.

\begin{deluxetable}{ll}
\tabletypesize{\scriptsize}
\tablewidth{0pt}
\tablecaption{Magnitudes of Host Galaxy} 
\tablehead{
\colhead{Band}  & \colhead{Magnitude (AB)}
}
\startdata
$B$ & 24.24$\pm$0.10 \citep{Levan2011}\\
$g$ & 23.67$\pm$0.19\\
F606W & 22.72$\pm$0.01\\
$r$ &  22.73$\pm$0.06\\
$i$ &  22.25$\pm$0.04\\
$Z$ & 22.16$\pm$0.06\\
$Y$ & 22.18$\pm$0.07\\
$J$ & 21.96$\pm$0.08\\
$H$ & 21.74$\pm$0.12\\
$K$ & 21.55$\pm$0.10\\
3.6$\mu$m & 21.70$\pm$0.07 (Including transient)\\
4.5$\mu$m & 21.57$\pm$0.08 (Including transient)
\enddata
\tablecomments{Magnitudes are corrected by Galactic extinction based on \citet{Schlafly2011}.}
\label{hosttab}
\end{deluxetable}

\begin{deluxetable*}{rl}
\tabletypesize{\scriptsize}
\tablewidth{0pt}
\tablecaption{Input Parameters for SED Fitting} 
\tablehead{
\colhead{Parameter} & \colhead{Value}
}
\startdata
$\tau$ ($e$-folding time scale of stellar population) & $6.5\leq$ $\log$[$\tau$/yr]  $\leq11.0$ with a step size of 0.1  \\
$t$ (age of stellar population) &  $8.0\leq$ $\log$[$t$/yr]  $\leq10.3$ with a step size of 0.1 \\
IMF (initial mass function) & \citet{Salpeter1955}  \\
Z (metallicity) & 0.004, 0.008, 0.020, 0.050 \\
Extinction law &  \citet{Calzetti2000} \\
$A_{V}$ ($V$-band attenuation for stellar population in magnitude) & $0.0\leq$ $A_{V}$  $\leq3.0$ with a step size of 0.1 
\enddata
\label{SEDtab}
\end{deluxetable*}

\begin{deluxetable}{rr}
\tabletypesize{\scriptsize}
\tablewidth{0pt}
\tablecaption{Best-fit Parameters from SED Fitting} 
\tablehead{
\colhead{Parameter} & \colhead{Value}
}
\startdata
Stellar mass [$\log(M_{\star}/M_{\odot})$] & $9.14^{+0.13}_{-0.10}$\\
\\
SFR [$M_{\odot}$/yr] & $0.03^{+0.28}_{-0.03}$\\
\\
Specific SFR [$\log(\rm{sSFR}/\rm{yr}^{-1})$] & $-10.62^{+0.90}_{-\infty}$\\
\\
$t$ [Gyr] & $0.63^{+0.95}_{-0.43}$\\
\\
$\tau$ [Gyr] & $0.10^{+0.24}_{-0.10}$\\
\\
$A_{V}$ & $0.00^{+0.97}_{-0.00}$\\  
\\
Z & $0.050^{+0.000}_{-0.046}$\\
\\
$\chi^2$  & 1.64
\enddata
\label{SED2tab}
\end{deluxetable}

 The right panel of Figure~\ref{lcfig} shows the light curves of the \emph{Spitzer} IRAC 3.6$\mu$m and 4.5$\mu$m bands. The magnitude changes in these bands are the more significant than those for the other optical/NIR bands. The fluxes from the transient component seem to be non-negligible even in the last epoch data  ($\Delta t = 333$ days), because the last IRAC epochs were still in the rapidly decreasing phase. Therefore, we consider the fluxes of the last epoch IRAC data to be the upper limit fluxes of the host galaxy.

 The multi-band magnitudes of the host galaxy of \emph{Swift} J1644+57 are shown in Table~\ref{hosttab}. We took the Galactic extinction of photometric data into account based on \citet{Schlafly2011}. We added the $B$-band photometric data from \citet{Levan2011} to expand the data points to the short wavelength band.

 The temporal change of the observed SED of \emph{Swift} J1644+57 are summarized in Figure~\ref{changefig}. The fluxes in redder bands show substantial changes with time. Meanwhile, bluer band fluxes are constant. This red feature of the transient has been suggested to be due to dust extinction. From previous studies, it is known that the hydrogen column density of the source of X-ray is large ($N_{\mathrm{H}}\sim10^{22}$cm$^{-2}$), meaning the line of sight to the SMBH has a very large extinction value ($A_{V}=4.5$ -- $10$) \citep{Bloom2011,Burrows2011,Levan2011,Shao2011,Saxton2012}.\\

\begin{figure}[!t]
\centering
\includegraphics[scale=0.296,angle=00]{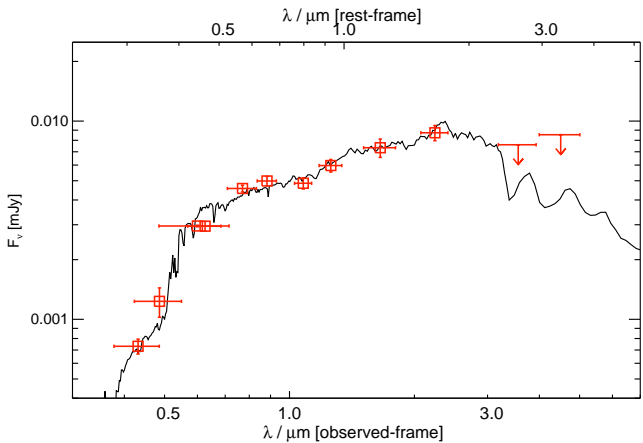}
	\caption{SED fitting result of the host galaxy of \emph{Swift} J1644+57. Plotted are the multi-band fluxes of the host galaxy in Table~\ref{hosttab}. Two \emph{Spitzer} data are the upper limit fluxes. The $\chi^2$ value for the fit is 1.64.
		\label{SEDfig}}
\end{figure}

\section{SED Fitting} \label{sec:SED}

We performed SED model fitting of the multi-band fluxes of host galaxy of \emph{Swift} J1644+57 in order to determine the properties of the host galaxy such as the stellar mass ($M_{\star}$) which is an important parameter for the $M_{\rm BH}$ estimation. We utilized the code Fitting and Assessment of Synthetic Templates \citep[FAST;\footnote{http://astro.berkeley.edu/$\sim$mariska/FAST.html}][]{Kriek2009}, which is a public SED fitting tool for the investigation of the galaxy properties using the photometric data ranging from UV to IR. The code is based on the IDL and fits of UV to IR stellar population templates to photometric data or galaxy spectra. FAST runs with the method of $\chi^2$ fitting and using stellar population grids to derive the best-fit model and its parameters. 

 We used the 2003 version of the Bruzual \& Charlot (BC03) model \citep{BC03} for the stellar population model. There are three initial mass functions (IMFs) available in the FAST \citep{Salpeter1955,Kroupa2001,Chabrier2003}. We chose the Salpeter IMF. To define the star formation history (SFH), we assumed an exponentially decreasing star formation rate (SFR). The stellar population was modeled with the $e$-folding time scales, $6.5\leq$ $\log$[$\tau$/yr]  $\leq11.0$ with a step size of 0.1 and ages of $8.0\leq$ $\log$[$t$/yr]  $\leq10.3$ with a step size of 0.1. We used several metallicity values such as Z = 0.004, 0.008, 0.02, and 0.05. The model SEDs were attenuated by dust, for which we used attenuation curves based on \citet{Calzetti2000}. We adopted $0.0\leq$ $A_{V}$  $\leq3.0$ with a step size of 0.1. All the input parameters for the SED fitting are summarized in Table~\ref{SEDtab}. 

 Figure~\ref{SEDfig} shows the SED fitting result. The two \emph{Spitzer} data were treated as the upper limit fluxes of the host galaxy. The estimated stellar mass of the host galaxy is $\log(M_{\star}/M_{\odot}) = 9.14^{+0.13}_{-0.10}$. The $e$-folding time scale is $\tau = 0.10^{+0.24}_{-0.10}$ Gyr and the age of the stellar population is $t = 0.63^{+0.95}_{-0.43}$ Gyr. The SFR of galaxy is $0.03^{+0.28}_{-0.03} \, M_{\odot}$/yr, and the specific SFR is $\log(\rm{sSFR}/\rm{yr}^{-1}) = -10.62^{+0.90}_{-\infty}$. \citet{Levan2011} derived SFR of $0.3$ -- $0.7 \, M_{\odot}$/yr from the H$\alpha$ and [O II] emission line luminosities. The value of $0.3 \, M_{\odot}$/yr from H$\alpha$ is consistent with our 1$\sigma$ upper limit. The SFR from [O II] line ($0.7 \, M_{\odot}$/yr) is about twice larger but the [O II] line based SFRs are known to be dependent on physical condition such as the reddening and metallicity \citep[e.g.,][]{Kewley2004}, and less reliable than H$\alpha$ based SFRs. Another possible cause of the discrepancy is the different timescales that are probed by different SFR indicators (emission line indicators probing recent star formation). \citet{Levan2011} estimated $E(B-V)_{gas}=-0.01 \pm 0.15$mag, i.e., no extinction using the intensity ratio of H$\alpha$ and H$\beta$ lines. This is consistent with our SED fitting result $A_{V}=0.00^{+0.97}_{-0.00}$. The $\chi^2$ value for the fit is 1.64.

 The host galaxy of \emph{Swift} J1644+57 is a low mass, low SFR galaxy with a low extinction. Also it seems to have experienced a rapid decline of SFR not very long ago. This fits in well with a recent suggestion by \citet{Arcavi2014} that host galaxies of tidal disruption events are E+A galaxies with $<1$ Gyr stellar population and low or no SFRs.

 The output parameters are given in Table~\ref{SED2tab}. The errors correspond to 1$\sigma$ confidence intervals derived from 100 times Monte Carlo simulations in which the input photometric data are changed according to their errors.
 
 We also tried the Chabrier IMF instead of the Salpeter IMF for the fit. The change of the IMF influenced to the stellar mass, decreasing the stellar mass by $\sim0.25$ dex.

 We also fitted the SED model with the \citet{Maraston2005} stellar population instead of the BC03 model. We set the input parameter ranges identical to the case of the BC03 stellar populations. The results were nearly identical to the BC03 result. 

 Our analysis of the host galaxy shows that the host galaxy is a bulge-dominated and nearly extinction free ($A_{V} \sim 0$ mag). On the other hand, the spectral properties of the nuclear transient suggests a high extinction ($A_{V} \sim 6$ mag). These two facts may appear contradictory, but we note that a significant amount of dust can be found easily in nuclear region of bulge-dominated galaxies when their nuclei are acitve. For example, hosts of luminous AGNs are mostly early-type, bulge-dominated galaxies \citep[e.g.,][]{Hong2015}, and such AGNs are known to contain a significant amount of dust in nuclear region as a form of hot or warm dusty torus \citep[e.g.,][]{Kim2015}. 
\\

\section{Discussion on Black Hole Mass} \label{sec:BH}
 Our results on the properties of the host galaxy of \emph{Swift} J1644+57 can be summarized as follows. It is a bulge-dominant galaxy (B/T=$0.83\pm0.03$). The mass of the host galaxy is somewhat low at $10^{9.14} \,M_{\odot}$, even though the galaxy is bulge-dominated. Now we estimate the mass of the SMBH that played the main role in the transient phenomenon.

 It is now generally accepted that the SMBHs ($10^6$ -- $10^{10} \,M_{\odot}$) reside in the bulges of all massive galaxies. Tight scaling relations have been derived between SMBH mass and several physical properties of the bulges (velocity dispersion, mass, luminosity, etc.) in many previous studies \citep{Magorrian1998,Ferrarese2000,Gebhardt2000,McLure2002,Marconi2003,Haring2004,Aller2007,Hopkins2007,Gultekin2009,Kormendy2009,Sani2011,Kormendy2013}. Some argue that ellipticals and classical bulges follow identical relations, while the pseudobulges follow a somewhat different relation with large scatter \citep{Hu2008,Kormendy2011,Sani2011,Kormendy2013}. We conclude from the best-fit galaxy models that the host galaxy of \emph{Swift} J1644+57 has a classical bulge, and have also found a minor possibility of the pseudobulge with B/T=$0.36$. For now, we consider only the best model, that is, the case of the host galaxy having a classical bulge and being bulge-dominant. 

\begin{figure}[!t]
\centering
\includegraphics[scale=0.175,angle=00]{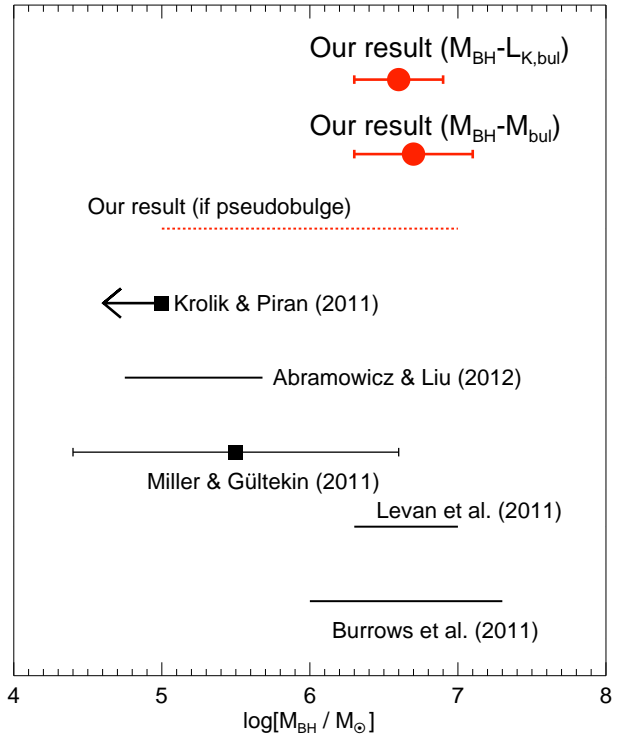}
	\caption{Results on the black hole mass from this work and previous studies. The red circles and dashed line represent our results, while the black squares and lines denote the results from previous studies. Error bars correspond to deviation of 1$\sigma$ and the arrow indicates the upper bound value.
		\label{BHfig}}
\end{figure}

 In order to esimate the central SMBH mass in the host galaxy of \emph{Swift} J1644+57, we used the scaling relation between $M_{\rm BH}$ and the stellar mass of the bulge ($M_{\star,\mathrm{bul}}$). We expect that a large part of the stellar mass derived in \S\ref{sec:SED} belongs to the bulge component. \citet{Sani2011} present the $M_{\mathrm{BH}}$ -- $M_{\star,\mathrm{bul}}$ relation, where $M_{\star,\mathrm{bul}}$ is directly obtained from the bulge luminosity ($L_{\mathrm{bul}}$) of \emph{Spitzer} 3.6$\mu$m and the calibrated $M_{\star,\mathrm{bul}}$ -- $L_{\mathrm{bul}}$ relation. They excluded pseudobulges when constructing the relation. The relation is 
\begin{equation}
	\mathrm{log}(M_\mathrm{BH}/M_{\odot})=\alpha+\beta\times[\mathrm{log}(M_\mathrm{\star,bul}/M_{\odot})-11],
\end{equation}
where $\alpha=8.16\pm0.06$, $\beta=0.79\pm0.08$, and the intrinsic scatter is $0.38\pm0.05$ . The estimated mass of the SMBH is $10^{6.7\pm0.4} \,M_{\odot}$ based on the stellar mass of the host galaxy and the above relation. If we consider the B/T=0.83 and assume that the mass-to-light ratio is constant in the bulge and disk, the stellar mass is decreased by $\sim0.1$ dex. It leads to a decrease in $M_{\mathrm{BH}}$ by $\sim0.1$ dex.

 The tight scaling relations between $M_{\mathrm{BH}}$ and host galaxy properties suggest a close link between the SMBH growth and the galaxy evolution. There may be a cosmic evolution of the scaling relations, for which there have been various studies \citep{Treu2004,McLure2006,Shields2006,Woo2006,Salviander2007,Treu2007,Woo2008,Jahnke2009,Bennert2010,Decarli2010,Merloni2010,Bennert2011}. The evolution of the scaling relations is still controversial in terms of the selection effects in the high redshift regime. Nevertheless, we can consider a case where the growth of $M_{\mathrm{BH}}$ happened ahead of the assembly of the stellar mass as suggested by some of these studies. \citet{Bennert2011} suggest the redshift evolution out to $z\sim2$, in the form of $M_{\mathrm{BH}}/M_{\star,\mathrm{bul}}\propto(1+z)^{1.96\pm0.55}$, based on 11 X-ray-selected broadline AGNs. Considering this evolution effect and the redshift of \emph{Swift} J1644+57 $z=0.35$, the mass of the SMBH could be larger by $\sim0.3$ dex.

 We also estimated $M_{\mathrm{BH}}$ through the $M_{\mathrm{BH}}$ -- $K$ band luminosity of bulge ($L_{K,\mathrm{bul}}$) relation in \citet{Kormendy2013}, assuming that most of the NIR fluxes come from the bulge. Their relation is much improved compared with previous studies in view of serveral things. They excluded galaxies with black hole monsters which have over-massive SMBHs despite having relatively small bulges and ellipticals. They also omitted galaxies with black hole masses that are measured based on the kinematics of ionized gas without taking line widths into account, since this method may yield underestimated masses. Galaxies in the process of merging generally have low-mass black holes for their luminosities. Thus they excluded these galaxies from the relation. They also did not include pseudobulges. The relation of \citet{Kormendy2013} is
\begin{equation}
	\mathrm{log}(M_\mathrm{BH}/10^9\,M_{\odot})=-\alpha-\beta\times(M_{K,\mathrm{bul}}+24.21),
\end{equation}
where the $\alpha$ is $0.265\pm0.050$, the $\beta$ is $0.488\pm0.033$, and intrinsic scatters is 0.30. $M_{K,\mathrm{bul}}$ is the $K$-band absolute magnitude of the bulge based on the photometric system of 2MASS. Using the best-fit SED derived earlier, we applied K-correction and evolution correction. We find $M_{K} = -19.84$ Vega mag including the evolutionary correction of 1.38 mag, which is derived by the difference in the $K$-band magnitude between 0.63 Gyr old population as in Table~\ref{SED2tab} and 4.5 Gyr old population, which is the age of this host galaxy in the local universe. Then, the mass of the SMBH is estimated to be $10^{6.6\pm0.3}\,M_{\odot}$. If we take B/T=0.83 into account and assume that this value is also applicable to the NIR bands, then $M_{\mathrm{BH}}$ decreases by $\sim0.1$ dex. If we use an $M_{K}$ value that has not been corrected for evolution, then $M_{\mathrm{BH}}$ becomes $\sim0.7$ dex larger.

 The tidal disruption of normal stars by a black hole is not possible for $M_\mathrm{BH}>10^8\,M_{\odot}$, since the tidal radius where the disruption can occur is located inside the Schwarzschild radius \citep{Rees1988,Cannizzo1990,Bloom2011}. Our results on the mass of the SMBH satisfy the condition for a tidal disruption event. 

Although we favor the model in which the host galaxy of \emph{Swift} J1644+57 is a classical bulge, our analysis shows that it could be a galaxy with a pseudobulge with B/T=0.36 (Table~\ref{galfittab}). If so, it is rather difficult to obtain an $M_{\rm BH}$ value, since the scaling relation is not well established for pseudobulges, especially in the low mass range of $M_{\star}\sim10^9\,M_{\odot}$ for the host galaxy. Several works have shown that the $M_{\rm{BH}}$ -- host galaxy scaling relations are weak or zero with a large scatter for pseudobulges. Over the $M_\mathrm{\star,bul}$ range of $10^{9.3}$ to $10^{10.5}\,M_{\odot}$ where such a relation has been studied, $M_{\rm BH}$ can have any value between $10^{6}$ to $10^{8}\,M_{\odot}$  \citep[Figure 21 of][]{Kormendy2013}. To reach down to $M_{\star,\mathrm{bul}} \sim 5 \times 10^{8}\,M_{\odot}$ as implied from the pseudobulge fit of our data, currently one can barely do so by relying on results from low mass AGNs \citep{Barth2005,Greene2008,Jiang2011,Xiao2011}. In such a case, an $M_{\rm BH}$ value between $10^{5}$ to $10^{6.3}\,M_{\odot}$ is possible \citep[Figure 32 of][]{Kormendy2013}. Overall, if the host galaxy harbors a pseudobulge, then we can only loosely constrain $M_{\rm BH}$ to have a value between $10^{5}$ to $10^{7}\,M_{\odot}$ considering our current poor knowledge of the $M_{\rm BH}$ value in pseudobulges. 

It is also known that a small fraction of pseudobulges have a S\'{e}rsic index of $n > 3$. The best example is Pox 52, for which $n\sim3.6$ -- $4.3$,  $M_{\rm BH} \sim2\times10^{5} \,M_{\odot}$, and $M_{\star} \sim10^9 \,M_{\odot}$ \citep{Barth2004,Thornton2008}. Therefore, even if we accept the S\'{e}rsic index of $n=3.43$ as the best-fit result, we need to keep this kind of caveat in mind.

 Figure~\ref{BHfig} shows our overall results on $M_{\mathrm{BH}}$ and the results from the previous studies we mentioned in \S\ref{sec:Intro}. It shows that our favorite results are compatible with the previous rough estimates from \citet{Burrows2011} and \citet{Levan2011}, who also used scaling relations. However, our results are improved compared to the previous results, by revealing that the host galaxy has a significant bulge component through a two-dimensional bulge + disk decomposition of the surface brightness profile, and removing the transient component in NIR light using a long-term light curve. The $M_{\rm BH}$ limit could be much looser (the dashed line) if the host galaxy harbors a pseudobulge. A critical test of the pseudobulge model would be to obtain a deep, high-resolution image to see how the surface brightness profile behaves at the outer region of the host galaxy. 
\\

\section{Summary}
 We investigated the host galaxy properties of tidal disruption event, \emph{Swift} J1644+57 through morphology analysis, light curve analysis, and SED fitting. We also estimated $M_{\mathrm{BH}}$ which played the main role of this phenomenon, through scaling relations.

 We decomposed the surface brightness profile of the host galaxy based on high-resolution \emph{HST} WFC3 images. We found that the host galaxy of \emph{Swift} J1644+57 is a bulge-dominated galaxy which is well described by a single S\'{e}rsic model with the S\'{e}rsic index, $n=3.43\pm0.05$. If we add a disk component, the bulge to total host galaxy flux ratio (B/T) is $0.83\pm0.03$, still indicating a bulge-dominant galaxy. We conclude that the host galaxy of \emph{Swift} J1644+57 has a classical bulge from the best-fit galaxy models, although we cannot completely exclude the possibility of this galaxy containing a pseudobulge with B/T=$0.36$.

 The NIR light curves enabled us to isolate the fluxes from the host galaxy after $\sim 500$ days following the dissipation of the X-ray flux. On the other hand, we found that there are no significant changes in the light curves of the short wavelength bands, supporting the red feature of the transient possibly being caused by severe dust extinction.

 We fit SEDs to the multi-band fluxes of the host galaxy which are derived in the light curve analysis.  The estimated stellar mass of the host galaxy is $\log(M_{\star}/M_{\odot}) = 9.14^{+0.13}_{-0.10}$. The $e$-folding time scale $\tau$ is $0.10^{+0.24}_{-0.10}$ Gyr and the age of stellar population is $0.63^{+0.95}_{-0.43}$ Gyr. The SFR of galaxy is $0.03^{+0.28}_{-0.03} \,M_{\odot}$/yr. In terms of the surface brightness profile and the stellar mass, this galaxy resembles M32, a small companion galaxy of M31. 

 We estimated the central $M_{\mathrm{BH}}$ through scaling relations. The mass of the SMBH is estimated to be $10^{6.7\pm0.4} \,M_{\odot}$ from $M_{\mathrm{BH}}$ -- $M_{\star,\mathrm{bul}}$ and $M_{\mathrm{BH}}$ -- $L_{\mathrm{bul}}$ relations for the $K$ band. However, the limit on $M_{\mathrm{BH}}$ can be much looser if the host galaxy has a pseudobulge. Future high-resolution, deep imaging should be able to unambiguosly distinguish the two possibilities.
\\

\acknowledgments 
This work was supported by the National Research Foundation of Korea (NRF) grant, No. 2008-0060544, funded by the Korea government (MSIP). We thank the observers who obtained the CQUEAN and UKIRT data that were used in our analysis. This paper includes the data taken at the McDonald Observatory of the University of Texas at Austin. At the time of the UKIRT observation, UKIRT was operated by the Joint Astronomy Centre on behalf of the Science and Technology Facilities Council of the U.K. This work is based in part on observations made with the Spitzer Space Telescope, which is operated by the Jet Propulsion Laboratory, California Institute of Technology under a contract with NASA. We acknowledge the use of public data from the Swift data archive. CP, TS, and NG acknowledge support from the NASA research grant, NNX10AF39G. MI gratefully acknowledges the hospitality and the support from the Korea Institute of Advanced Study where part of this was carried out.
\\


\begin{thebibliography}

\bibitem[Abramowicz 
\& Liu(2012)]{al2012} Abramowicz, M.~A., \& Liu, F.~K.\ 2012, \aap, 548, A3

\bibitem[Aller 
\& Richstone(2007)]{Aller2007} Aller, M.~C., \& Richstone, D.~O.\ 2007, \apj, 665, 120 

\bibitem[Arcavi et al.(2014)]{Arcavi2014} Arcavi, I., Gal-Yam, A., 
Sullivan, M., et al.\ 2014, \apj, 793, 38 

\bibitem[Barth et al.(2004)]{Barth2004} Barth, A.~J., Ho, L.~C., 
Rutledge, R.~E., \& Sargent, W.~L.~W.\ 2004, \apj, 607, 90 

\bibitem[Barth et al.(2005)]{Barth2005} Barth, A.~J., Greene, 
J.~E., \& Ho, L.~C.\ 2005, \apjl, 619, L151 

\bibitem[Bennert et al.(2011)]{Bennert2011} Bennert, V.~N., Auger, 
M.~W., Treu, T., Woo, J.-H., \& Malkan, M.~A.\ 2011, \apj, 742, 107 

\bibitem[Bennert et al.(2010)]{Bennert2010} Bennert, V.~N., Treu, 
T., Woo, J.-H., et al.\ 2010, \apj, 708, 1507 

\bibitem[Bertin 
\& Arnouts(1996)]{Bertin1996} Bertin, E., \& Arnouts, S.\ 1996, \aaps, 117, 393 

\bibitem[Bloom et al.(2011)]{Bloom2011} Bloom, J.~S., Giannios, 
D., Metzger, B.~D., et al.\ 2011, Science, 333, 203 

\bibitem[Bruzual 
\& Charlot(2003)]{BC03} Bruzual, G., \& Charlot, S.\ 2003, \mnras, 344, 1000 

\bibitem[Burrows et al.(2011)]{Burrows2011} Burrows, D.~N., Kennea, 
J.~A., Ghisellini, G., et al.\ 2011, \nat, 476, 421 

\bibitem[Calzetti et al.(2000)]{Calzetti2000} Calzetti, D., Armus, 
L., Bohlin, R.~C., et al.\ 2000, \apj, 533, 682 

\bibitem[Cannizzo et al.(1990)]{Cannizzo1990} Cannizzo, J.~K., Lee, 
H.~M., \& Goodman, J.\ 1990, \apj, 351, 38 

\bibitem[Chabrier(2003)]{Chabrier2003} Chabrier, G.\ 2003, \pasp, 
115, 763 

\bibitem[Decarli et al.(2010)]{Decarli2010} Decarli, R., Falomo, 
R., Treves, A., et al.\ 2010, \mnras, 402, 2453 

\bibitem[de Vaucouleurs(1948)]{deVa1948} de Vaucouleurs, G.\ 
1948, Annales d'Astrophysique, 11, 247

\bibitem[Ferrarese 
\& Merritt(2000)]{Ferrarese2000} Ferrarese, L., \& Merritt, D.\ 2000, \apjl, 539, L9 

\bibitem[Fisher 
\& Drory(2008)]{Fisher2008} Fisher, D.~B., \& Drory, N.\ 2008, \aj, 136, 773 

\bibitem[Fisher 
\& Drory(2010)]{Fisher2010} Fisher, D.~B., \& Drory, N.\ 2010, \apj, 716, 942 

\bibitem[Gebhardt et al.(2000)]{Gebhardt2000} Gebhardt, K., Bender, 
R., Bower, G., et al.\ 2000, \apjl, 539, L13 

\bibitem[Greene et al.(2008)]{Greene2008} Greene, J.~E., Ho, 
L.~C., \& Barth, A.~J.\ 2008, \apj, 688, 159 

\bibitem[G{\"u}ltekin et al.(2009)]{Gultekin2009} G{\"u}ltekin, K., 
Richstone, D.~O., Gebhardt, K., et al.\ 2009, \apj, 698, 198 

\bibitem[H{\"a}ring 
\& Rix(2004)]{Haring2004} H{\"a}ring, N., \& Rix, H.-W.\ 2004, \apjl, 604, L89 

\bibitem[Hong et al.(2015)]{Hong2015} Hong, J., Im, M., Kim, M., \& Ho, L.~C.\ 2015, \apj, 804, 34 

\bibitem[Hopkins et al.(2007)]{Hopkins2007} Hopkins, P.~F., 
Hernquist, L., Cox, T.~J., Robertson, B., 
\& Krause, E.\ 2007, \apj, 669, 67 

\bibitem[Hu(2008)]{Hu2008} Hu, J.\ 2008, \mnras, 386, 2242 

\bibitem[Jahnke et al.(2009)]{Jahnke2009} Jahnke, K., Bongiorno, 
A., Brusa, M., et al.\ 2009, \apjl, 706, L215 

\bibitem[Jiang et al.(2011)]{Jiang2011} Jiang, Y.-F., Greene, 
J.~E., \& Ho, L.~C.\ 2011, \apjl, 737, L45 

 \bibitem[Kewley et al.(2004)]{Kewley2004} Kewley, L.~J., Geller, 
M.~J., \& Jansen, R.~A.\ 2004, \aj, 127, 2002 

\bibitem[Kim et al.(2011)]{KimE2011} Kim, E., Park, W.-K., 
Jeong, H., et al.\ 2011, Journal of Korean Astronomical Society, 44, 115 

\bibitem[Kim et al.(2015)]{Kim2015} Kim, D., Im, M., Kim, J.~H., et al.\ 2015, \apjs, 216, 17 

\bibitem[Kormendy 
\& Bender(2009)]{Kormendy2009} Kormendy, J., \& Bender, R.\ 2009, \apjl, 691, L142 

\bibitem[Kormendy et al.(2011)]{Kormendy2011} Kormendy, J., Bender, 
R., \& Cornell, M.~E.\ 2011, \nat, 469, 374 

\bibitem[Kormendy 
\& Ho(2013)]{Kormendy2013} Kormendy, J., \& Ho, L.~C.\ 2013, \araa, 51, 511 

\bibitem[Kriek et al.(2009)]{Kriek2009} Kriek, M., van Dokkum, 
P.~G., Labb{\'e}, I., et al.\ 2009, \apj, 700, 221 

\bibitem[Krolik 
\& Piran(2011)]{kp2011} Krolik, J.~H., \& Piran, T.\ 2011, \apj, 743, 134 

\bibitem[Kroupa(2001)]{Kroupa2001} Kroupa, P.\ 2001, \mnras, 322, 
231 

\bibitem[Lee et al.(2010)]{Lee2010} Lee, I., Im, M., 
\& Urata, Y.\ 2010, Journal of Korean Astronomical Society, 43, 95 


\bibitem[Levan et al.(2011)]{Levan2011} Levan, A.~J., Tanvir, 
N.~R., Cenko, S.~B., et al.\ 2011, Science, 333, 199 

\bibitem[Levan 
\& Tanvir(2012)]{Levan2012} Levan, A.~J., \& Tanvir, N.~R.\ 2012, The Astronomer's Telegram, 4610, 1 

\bibitem[Lim et al.(2013)]{Lim2013} Lim, J., Chang, S., Pak, 
S., et al.\ 2013, Journal of Korean Astronomical Society, 46, 161 

\bibitem[Magorrian et al.(1998)]{Magorrian1998} Magorrian, J., 
Tremaine, S., Richstone, D., et al.\ 1998, \aj, 115, 2285 

\bibitem[Maraston(2005)]{Maraston2005} Maraston, C.\ 2005, \mnras, 
362, 799 

\bibitem[Marconi 
\& Hunt(2003)]{Marconi2003} Marconi, A., \& Hunt, L.~K.\ 2003, \apjl, 589, L21 

\bibitem[McLure 
\& Dunlop(2002)]{McLure2002} McLure, R.~J., \& Dunlop, J.~S.\ 2002, \mnras, 331, 795 

\bibitem[McLure et al.(2006)]{McLure2006} McLure, R.~J., Jarvis, 
M.~J., Targett, T.~A., Dunlop, J.~S., 
\& Best, P.~N.\ 2006, \mnras, 368, 1395 

\bibitem[Merloni et al.(2010)]{Merloni2010} Merloni, A., Bongiorno, 
A., Bolzonella, M., et al.\ 2010, \apj, 708, 137 

\bibitem[{{Miller} \& {G{\"u}ltekin}(2011)}]{mg2011}
{Miller}, J.~M. and {G{\"u}ltekin}, K. 2011, \apjl, 738, L13

\bibitem[Park et al.(2012)]{Park2012} Park, W.-K., Pak, S., Im, 
M., et al.\ 2012, \pasp, 124, 839 

\bibitem[Peng et al.(2010)]{Peng2010} Peng, C.~Y., Ho, L.~C., 
Impey, C.~D., \& Rix, H.-W.\ 2010, \aj, 139, 2097 

\bibitem[Rees(1988)]{Rees1988} Rees, M.~J.\ 1988, \nat, 333, 523 

\bibitem[Salpeter(1955)]{Salpeter1955} Salpeter, E.~E.\ 1955, \apj, 
121, 161 

\bibitem[Salviander et al.(2007)]{Salviander2007} Salviander, S., 
Shields, G.~A., Gebhardt, K., \& Bonning, E.~W.\ 2007, \apj, 662, 131 

\bibitem[Sani et al.(2011)]{Sani2011} Sani, E., Marconi, A., 
Hunt, L.~K., \& Risaliti, G.\ 2011, \mnras, 413, 1479 

\bibitem[Saxton et al.(2012)]{Saxton2012} Saxton, C.~J., Soria, 
R., Wu, K., \& Kuin, N.~P.~M.\ 2012, \mnras, 422, 1625 

\bibitem[Schlafly 
\& Finkbeiner(2011)]{Schlafly2011} Schlafly, E.~F., \& Finkbeiner, D.~P.\ 2011, \apj, 737, 103 

\bibitem[Sersic(1968)]{Sersic1968} Sersic, J.~L.\ 1968, Atlas de Galaxias Australes (Cordoba, Argentina: Observatorio Astronomico, Univ. de Cordoba)

\bibitem[Shao et al.(2011)]{Shao2011} Shao, L., Zhang, F.-W., 
Fan, Y.-Z., \& Wei, D.-M.\ 2011, \apjl, 734, L33 

\bibitem[Shields et al.(2006)]{Shields2006} Shields, G.~A., 
Menezes, K.~L., Massart, C.~A., \& Vanden Bout, P.\ 2006, \apj, 641, 683 

\bibitem[Thornton et al.(2008)]{Thornton2008} Thornton, C.~E., 
Barth, A.~J., Ho, L.~C., Rutledge, R.~E., 
\& Greene, J.~E.\ 2008, \apj, 686, 892 

\bibitem[Treu et al.(2004)]{Treu2004} Treu, T., Malkan, M.~A., 
\& Blandford, R.~D.\ 2004, \apjl, 615, L97 

\bibitem[Treu et al.(2007)]{Treu2007} Treu, T., Woo, J.-H., 
Malkan, M.~A., \& Blandford, R.~D.\ 2007, \apj, 667, 117 

\bibitem[Woo et al.(2006)]{Woo2006} Woo, J.-H., Treu, T., 
Malkan, M.~A., \& Blandford, R.~D.\ 2006, \apj, 645, 900 

\bibitem[Woo et al.(2008)]{Woo2008} Woo, J.-H., Treu, T., 
Malkan, M.~A., \& Blandford, R.~D.\ 2008, \apj, 681, 925 

\bibitem[Xiao et al.(2011)]{Xiao2011} Xiao, T., Barth, A.~J., 
Greene, J.~E., et al.\ 2011, \apj, 739, 28 

\bibitem[Zauderer et al.(2013)]{Zauderer2013} Zauderer, B.~A., 
Berger, E., Margutti, R., et al.\ 2013, \apj, 767, 152 

\bibitem[Zauderer et al.(2011)]{Zauderer2011} Zauderer, B.~A., 
Berger, E., Soderberg, A.~M., et al.\ 2011, \nat, 476, 425 

\end{thebibliography}
\end{document}